# Transdisciplinary Multi Modal Approach to Knowledge

Johanna Casado[1], [1]Beatriz García[2] and Natasha Bertaina[1]

[1]*ITeDA (CNEA, CONICET, UNSAM), Universidad de Mendoza, Argentina*
[2]*ITeDA (CNEA, CONICET, UNSAM), UTN-FRM, Argentina*

*[johanna.casado@um.edu.ar](mailto:johanna.casado@um.edu.ar), [n.bertaina@alumno.um.edu.ar](mailto:n.bertaina@alumno.um.edu.ar),*
*[beatriz.garcia@iteda.cnea.gov.ar](mailto:beatriz.garcia@iteda.cnea.gov.ar)*

*Abstract*

*The muti-modal or multi-sensorial perception of nature is presented in this article as part of research devoted to inclusive tools developed in the framework of User Centered Design. This proposal shows that it is possible to work in a transdisciplinary way, establishing feedback not only between designers and final users, but also between humans and computers, to reduce errors and co-design the resources according to personal needs. As part of the present research, we present the basis for a new accessible software, sonoUno, which was designed with the user in mind from the beginning, and we propose a training activity to enhance the user's capacities, expand the detection of different kinds of natural signals, and improve the comprehension of the Human Computer Interfaces, opening new windows to the sciences for diverse populations, not only in education and outreach but also in research. Some examples of the exploitation of these new devices and tools are also presented.*

## 1. Introduction

Multimodal perception (use of more than one sense) has proven to be a valid approach for exploring and understanding complex scientific data. For more than a decade, we have investigated the effectiveness of the use of more than one sensory pathway for the study of nature, in a framework  that is not only multi sensorial, but also interdisciplinary, exploring existing computer resources while, at the same time, designing those that improve:

- the use of tools for data sonification,
- training users to improve their skills , and
- the possibility of inclusion and equity in the professional scientific world.

---

[1] Corresponding author

According to the World Health Organization (2011), about 15% of the world's population has some type of disability. Disabilities can be the product of various factors, but some of them have to do with living conditions and the aging of the world population, which confronts us with the possibility that all inhabitants develop some type of disability during their active lives.

Disability is not an attribute of the person, it is the result of the environment. Consequently, the barriers that arise from the interaction between people with disabilities and society prevent them from participating equally, fully, and effectively in any activity that they propose, not only at a professional level but also at school and outreach activities.

In generic terms, people with disabilities see their chances of doing any job they set out to do drastically reduced. In this sense, assistive technologies allow the removal of certain barriers and the integration of people with disabilities into active society.

Understanding transdisciplinarity as a cognitive scheme that allows "crossing" disciplines, or the space between two dimensions where a constant flow of information is evident, we can understand that the research strategy addressed in this paper really crosses disciplinary boundaries. Here an approach is proposed directly linked to the design of software for astrophysical data sonification (in tables or images) (combining programming, human perception, user centred design, accessibility, between other characteristics), effective in many other investigations, including those related to the perception and learning curves of diverse populations. This approach has allowed input and analysis in the scientific and non-scientific interested communities and has allowed the challenge of continuous improvement and user-centered design to be addressed in a systematic way, which ensures traceability. This way of approaching a currently important topic, such as the use of more than one sense for data analysis, includes initiatives that support the development of capacities required for the formulation and successful transdisciplinary implementation of research actions.

## 2. New windows into the nature

In so-called interdisciplinarity, there are complex intermediate (fractional) levels of interaction between various disciplines that address the same problem or object of study; it is here that high levels of conceptualization are required. The practical dimension can only be apprehended from the interpretation at various levels of reality and from constructive cognitive schemes (non-linear, multi-causal and even random) (Morin, 1990). Such is the case with so-called User-Centered Design.

All printed material, including text, illustrations, and charts, must keep astronomy (and sciences in general) as a professional field, only benefiting those that can cope with the performance styles currently available. But, evidently, this situation is not restricted only to the sciences; the approach to the environment based on vision, in science, education, and everyday life, is part of the limitation of the human being's potential.

Every day we use the term technology to refer to new things, sometimes strange, exotic, and even intimidating. However, technology refers to systematic applications of knowledge to create artifacts, materials, and even procedures. We might think that modern times are characterized by substantial progress in what we call technology: automation. There could be some objections to the "taking control" of human activities by this type of technology; however, it is the people who finally decide or take control of them since many times these technologies fail. In this sense, what is normally called a human-computer interface is a proposal in permanent development, where there is no way to advance without permanent feedback and updating, which often means the impossibility of having what would normally constitute a "market product", such as a smart tv or a stereo. The technologies that seek to ensure inclusion find that the most important thing is the human factor, and that is where inter, intra, and transdisciplinarity play a fundamental role.

Once it is understood that a person is the center of this type of development by maintaining knowledge of the context through what Donald Norman (2010) calls "awareness of the situation", it is easy to imagine that control, evaluation of the situation, and intervention in case of failure remain in your hands. For this to happen, the person must be permanently informed, and any user-centered design must be promoted, ensuring a careful relationship between the human and the machine.

We can then ask ourselves if, eventually, just as the user improves the performance of, for example, certain software (a screen reader, a sonification tool), machines could increase the capabilities of the person by providing tools and methods that enhance their skills. We are not talking about automating a task, but about increasing capacity.

## 3. Assistive Technologies

When people interact with each other, a series of natural protocols, signals, or spontaneous tactics are followed. When communicating with machines, it is also possible to establish certain rules. In this sense, we could adhere to the idea that to shorten the Human-Machine communication path, in our case, the computer, there are certain rules for human designers of "intelligent" machines (Norman, 2010):

>    1. Offer rich, complex, and natural signals.
>    2. Be predictable.
>    3. Offer a good conceptual model.
>    4. Make the result understandable.
>    5. Offer continuous awareness that doesn't bother.
>    6. Use natural correspondences to make the interface efficient and comprehensive.

When introducing intelligent systems, we are facing a great challenge since there is a lack of common ground between humans and computers. For interaction to be effective, it must be predictable and understandable. People must understand what is happening and naturally interact with machines, while machine states must also be continuously and efficiently generated.

It is also interesting to summarize what the machines propose to improve the relationship with humans, because this is where a model will emerge that is eventually mutually beneficial and should be targeted for the development of these new technologies:

>    1. Keep things as simple as possible

2. Provide people with a conceptual model
3. Explain the reasons for each action
4. Make sure people think they are the ones in control
5. Constantly reassure human users
6. Never label human behavior as a "mistake"

We are probably still far from these principles, but a user-centered design seeks precisely these same objectives. Taking into consideration all this technology and the principle of human computer interaction, assistive technology (AT) is particularly important to achieve inclusion.

Nowadays, the focus on achieving equity between people is centred on reinforcing their capabilities, where assistive technologies play an important role. Today, the majority of people use these technologies to access computer resources, and they are the ones who must adapt to the technology and not the opposite. Following the principles of the Convention on the Rights of Persons with Disabilities and all the actual efforts to reach equity for functionally diverse people, in conjunction with electronics and computational advances (smaller devices and artificial intelligence), may lead to the future of assistive technology where the tool reinforces at each moment the person's capabilities.

Assistive technology can help disabled people better access education, work, or development in their daily lives. Specific examples of these tools include:

- SOFTWARE and DEVICES for computer: screen readers, braille displays, braille keyboard, pictograms, voice recognition and screen magnifiers.
- EQUIPMENT: includes mobility aids, such as a wheelchair.

On the other hand, software like SonoUno (Casado et al, 2022b), which will be described in this article, uses assistive technology to assure equity between its users and produce a multimodal display of scientific information. It is a novel approach, where the talk of bringing equity and accessibility to the scientific field began. One of the SonoUno principles was to ensure good communication between the software and the assistive technology, making it possible for people to choose the AT and the sensorial modality of exploration. Some examples of using sonoUno tools with screen readers in Windows and Mac operating systems could be found in the Screen Readers list inside the sonoUno YouTube Channel[2].

The application of software for data sonification, accompanied by the development of specific training that allows its scientific use and which, in some cases, is complemented by the inclusion of peripherals, shows that efforts to remove the social systemic barriers that prevent real equality and inclusion of people by accepting diversity are possible.

### 3.1 Considerations for a new approach, a new software

This proposal is devoted to comprehensively reviewing the assistive technologies based on a design centered on the final user, but in the presentation we will focus on the sonoUno software development, taking into account the trans-disciplinary approach.

To analyze a new tool, we propose the following steps:

- Use of resources and techniques tested.
- Evaluation of new resources, ideas and revision: feedback with users.
- Functional diverse scientists as advisors.
- Study of Impact - different audiences, including peer reviews.

sonoUno is a new user-centered software to produce audio-visual outputs from 1D, 2D and 3D data; it integrates multidisciplinary and interdisciplinary scientists, both blind and sighted, and addresses the topic of accessibility to scientific data. This development follows all the steps mentioned above.

The approach is multisensorial, which means that it presents information in different sensorial styles: vision (with images and the traditional GUI), audition (sonification of the data and description of the GUI with screen readers), and tactile models (with 3D prints, for example).

After some years since the first version of sonoUno was released, a platform to support cross-reflection between citizens and scientists, but also between scientists in different disciplines, seems to be the best way to communicate the purposes of the research.

---

[2] https://youtube.com/playlist?list=PLKiFa5ln9IgEv9zmsLyv8gtVIlGshKHSq

In this sense, the steps to reach the objectives can be summarized in three main items:

> 1. A human-computer interface suitable for the access, sonification, analysis and collection of scientific data.
>
> 2. Test the efficiency, usefulness and effectiveness of the resource in different cultural environments.
>
> 3. Develop the paradigm for training researchers and interested citizen scientists to start using new techniques.

**4. Human computer interface**

For the software preparation, the original idea was also based on transdisciplinarity, and because of this, the design was thought of as a modular scheme that could be used part by part or as a whole. Five modules were defined:

- Data input: manage the request and convert the files into arrays which could be understood by Python language;
- Data output: the requirement is received, the location is demanded to the user, and then the file is created and stored;
- Sound module: conduct the sonification process and take care the sonification settings while incorporating;
- Math operations: collect the data and the operation mark by the user and return the result of this action;
- Graphics User Interface (GUI) module: contain the graphic user interface design, and the definition of each button task;
- Core module: use all the modules before, to link each button task with the method that takes care of it.

Following this, the functionalities were written on some pieces of paper, **Figure 1** presents the development design as it was discussed in preliminary meetings with specialists in different disciplines (astronomers, educators, bio-engineers, able and disabled people). Note the disposition of different resources in the software and the relation between each neighboring action: the labels correspond to functions (buttons) on the screen of the computer, and the distribution is connected with the simplest learning curve, even in the case of screen reader tools.

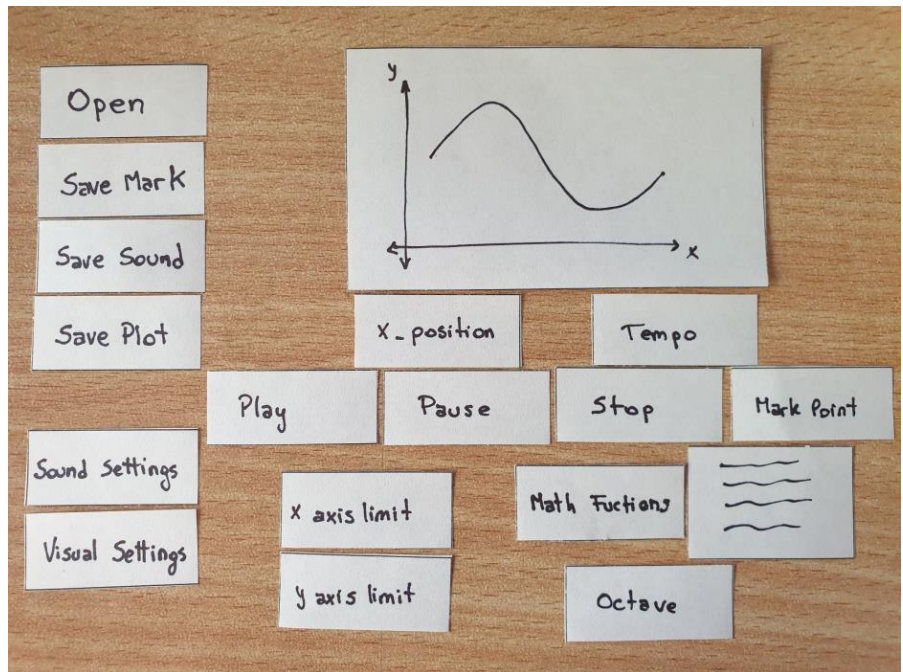

**Figure 1**: Preliminary Design for the sonoUno Functionalities Screen Deployment

The final graphical design is shown in **Figure 2**, where the modular design is highlighted at left and the final GUI at right (with the functionalities grouped by its principal goal). The categorization of tasks was based on the user and how a person reads an interface, so the groups were: data display (**Figure 2,** yellow box), which contains the buttons to manage the data display and a second section with a grid that shows the data and some additional options like selecting which column plot); data operations (pink box in **Figure 2**, under the data display panel, which group the mathematical functions and cut limit sliders); data configurations (**Figure 2,** blue box at the left of the interface), which contain all the plot and sound configurations); and input/output (**Figure 2,** red box at the top left of the interface), which enclose the input/output actions).

This special characteristic permits enlarging the usability of the software, abording new functionalities module by module, and in this way, producing specific sonorization scripts for specific applications or using the GUI interface for special plot deployments. The flexibility in the design is also part of the transdisciplinary view and a door for co-creation: different users and designers can contribute to improving the tool according to their needs, the complexity of the research, their level of education, or their cultural environments.

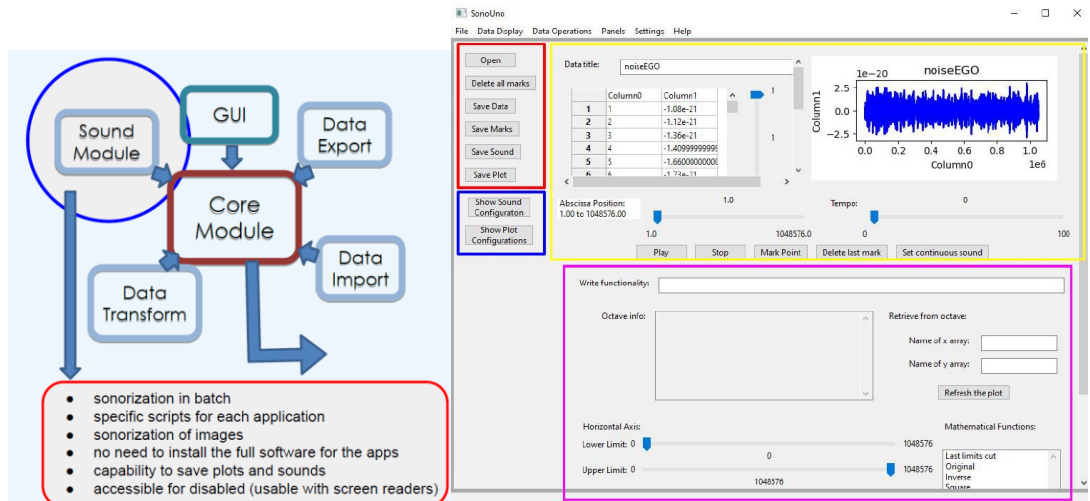

**Figure 2:** Modular design. The sound module can be modified according the user needs (left); final GUI of sonoUno for the desktop version (right)

The Human Computer Interface was prepared following the open source philosophy in Python, at the beginning for a desktop PC or netbook, for Windows, Mac, and Linux operating systems. After the exchange with the users and some difficulties with the installation, mainly connected with the changes in Python versions and the updates of the operating systems, the preparation of a web version starts. This web version is based on Java tools and the Angular framework, and also adopts the ARIA protocol to ensure accessibility and the use of the GUI with screen readers.

To maintain the open access quality, all the resources, including the source code, are available at the GitHub[3] and Zenodo[4] platforms, and part of the demos available at the sonoUno YouTube Channel[5].

A sign of inclusivity is not only the possibility to access the development in more than one sense but also to have it in different languages. The software is available for a moment in English and Spanish. Moreover, the sonoUno web page[6] presents a plug-in to show the content in seven languages (English, Spanish, Portuguese, French, German, Greek and Italian).

---

[3] https://github.com/sonoUnoTeam
[4] https://zenodo.org/communities/sonouno/
[5] https://www.youtube.com/channel/UCLVj2IiGOs-L0uXKZ8-8BLA
[6] https://www.sonouno.org.ar/

## 5. Training for usability

It is essential to consider that translation between "different languages" is not enough to assure equity and inclusion; it is also needed for the appropriation of knowledge. In that sense, it is necessary to keep in mind that perception (in the way the human being comprehends the stimulus received by their senses) is a process that requires a great deal of mental processing, which provides the means by which one's concept of the environment is created and which helps each person learn and interact with it. The compilation of previous studies throughout history has led to the conclusion that auditory performance improves when combined with visual stimuli and vice versa. This fact, related to the need to learn how to use sound as a reinforcement in data analysis (expressed by sonoUno users), leads the group to explore training sessions with multisensorial deployment. Bertaina Lucero et. al. (2022) explain the absence of training devoted to sonification and describe a training session conducted with different data sets.

During 2022 and 2023, different training sessions were designed that combined visual and auditory stimuli with the goal of teaching the use of more than one sense in data analysis and improving the user's performance. **Figure 3** shows one example of the training interface, from the block screen to the feedback after the user's answer.

The training sessions were modified as they were tested in different instances by users. During these experiences, in addition to data on the performance of the participants, suggestions for improving the training were extracted. Among all statements, the most outstanding were:

- Take a training course that increases in complexity.
- Allow reproducing the sound together with the visual display of the signal (in cases where each stimulus is reproduced alone).
- Include sound feedback that accompanies the one displayed visually.
- Offer the ability to skip certain introductory windows.

Among some of the comments that were made regarding the sonification, the desire to know more about the sound was mentioned, which highlights the need for a study focused on perception and more training courses. Respecting the sonification system, in said meeting, comments were received that evidenced the user's interest in more information about sonification, which highlights the need for a study focused on perception and more training courses (Lucero, 2023).

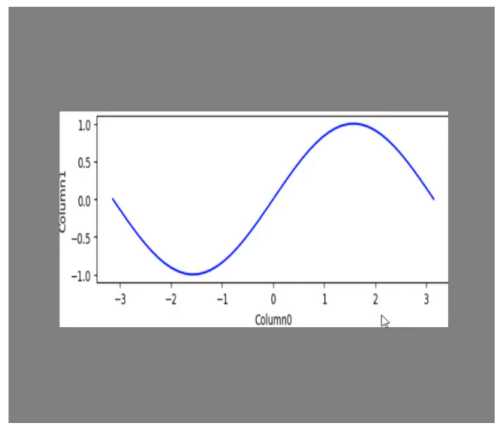

**Figure 3:** Visual display presented in a training session. A) Presentation of the first block of signals displayed in one of the designed training sessions. B) Signal displayed, which was accompanied by its sound[7]. C) Response options displayed for the participant. D) Feedback delivered to the participant immediately after the answer.

The key outcome from the last training session was that 77% of the participants were able to know the meaning of multisensory data analysis after the training. This is not minor since all the attendees (13 in total) had never used sonification, and 40% of them planned to incorporate the tool as part of their professional activities. Regarding the positive detections in the characteristics of the data after training, between 67 and 90% of assistants obtain positive results. All the training sessions support the importance and need for education on such a novel topic as the multisensorial display of scientific data.

---

[7] https://www.sonouno.org.ar/wp-content/uploads/sites/9/2023/05/sinusoidal_sound.wav

The statistic presented is relevant to this contribution because it reinforces the need for a user centred design and the importance of a multi sensorial approach. The user has been present from the beginning in each development by the sonoUno team, which results in usable tools and a new and inclusive approach to scientific data.

## 6. Impact of the approach and Exploitation

As part of the objectives of this work, the use of the tool in different environments and with different audiences was developed. Groups in Spain, USA and Argentina used the sonoUno to produce specific output with interesting results related not only to science and discoveries but also to the impact on the participants and the co-creating possibilities that open source software allows.

We selected a few examples of the exploitation of the sonification tool (**Figure 4**):

- The work on variable stars by Carlos Morales Socorro[8], in Spain, permitted after a very interesting training program, the discovery of a variable star by a blind student, using sonification, for the first time in the history[9].
- The group at Harvard, responsible for the project "Sensing the dynamic Universe"[10] (Díaz-Merced et al, 2020), generated a series of visual-sound material for variable stars, with very interesting derivation and future work possibilities.
- The analysis of a set of open data from Pierre Auger Observatory (The Pierre Auger Collaboration, 2020), showed the new window to discrete events sonorization.

t is evident that the human-computer interfaces (devices, software) to produce new output results can produce a virtuous relationship.

---

[8] https://www.flickr.com/photos/64618322@N07/albums/72177720299119160
[9] https://astronomiayeducacion.org/taller-2-de-sonificacion-descubriendo-el-universo/?cn-reloaded=1
[10] https://lweb.cfa.harvard.edu/sdu/

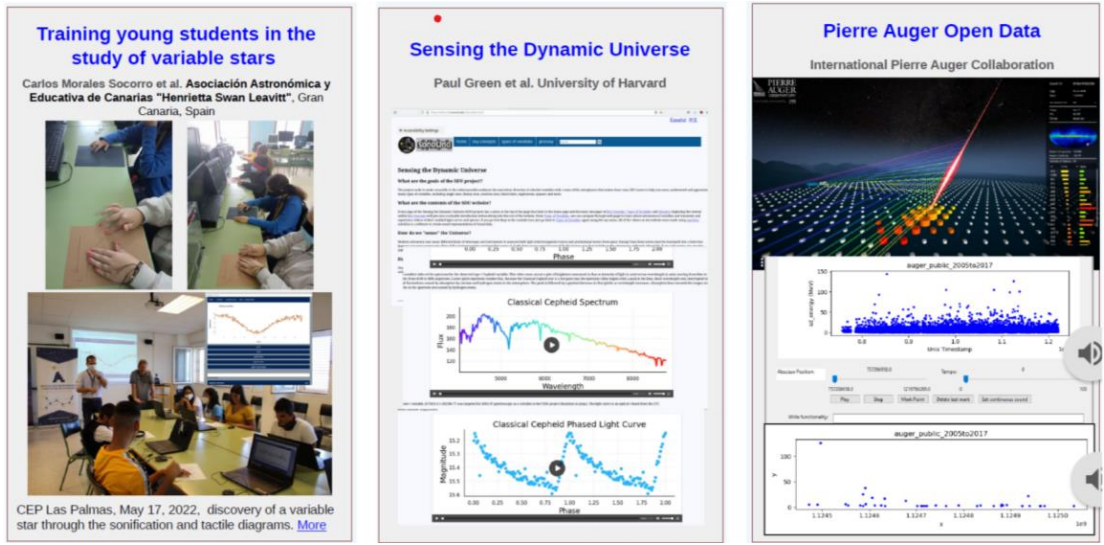

**Figure 4:** Different uses of sonoUno. Study and detection of variable stars (left); Sensing the dynamic universe: light curves and spectra (center); high energy cosmic rays at Pierre Auger Observatory (right)

The way to face the topic of human-computer interaction is changing. New methods and approaches propose resources called "quiet technology" or "environmental technology" (Waiser, 1991), which are really interesting things in which the computers and their tools will not be as stressful and frustrating that they can be in the present. In this line of thought, the UCD has, as one of its objectives, designed systems that accept the emerging problems, products of a bad relationship between humans and machines, and not only try to correct them but also establish a certain level of confidence between both.

However, these new ideas need to be supported by the entire community. We need to suggest nations carry out assessments about the gap between what organizations report as achieved and the quality of participation of the disabled and systematise a report format where there is evidence about the impact of the UCD in the development of HCI and databases. But at the same time, the nations, institutions, and funding agencies must be ready to stimulate the inclusion of topics and people with disabilities and other diversities in the projects and to include tools tested for people with different learning styles at the high level of government decisions.

## 7. Conclusions

In recent years, there has been evidence of an exponential growth of programs and techniques of data sonorization, mainly in astronomy and astrophysics, marking the pulse of these times in that the multimodal approach to the study of nature is emerging as a topic of high impact in the scientific community and the social field in general.

The framework presented in this article focused on a permanent commitment to inclusion and user-centered design. We have shown that a transdisciplinary approach is possible in this type of proposal.

Regarding sound techniques, although there are some antecedents before the 1990s, most of the developments and publications that evidence their use in science and popularization are after that date. According to Zanella et al. (2022), the last five years equaled, in percentage of new developments, the past 20 years. This shows the need that exists on the part of scientists (and society) for new data display techniques, reinforcing the premise that a multisensory display allows for better data analysis, but this is also connected with the demand for better tools among the so-called "intelligent" or "assistive" technologies.

Bearing in mind that a multimodal deployment also allows access to data for people with different sensory and learning styles, new "visualization" techniques, and the development of user-centered and accessible tools from the beginning, it seems to be the most appropriate way to obtain successful results in what concerns human-computer interfaces and the new exchange between users and machines.

## 8. Acknowledgments

The authors want to give their thanks to the people who tested the tool, participated in the training activities, and shared their experiences, ideas, suggestions, and comments. The contributions and feedback of Carlos Morales Socorro, Richard Green, Poshak Gandhi, Gaston Jaren, students at the University of Mendoza and ITeDA, and many volunteers at the Focus Groups and beyond testing the software are deeply appreciated.

We especially want to recognize the work of the referee, Poshak Gandhi, from the University of Southampton, UK; their comments and suggestions contributed to improving this article. The reading by Gaston Jaren, from the University of Mendoza, who reviewed for the expressive form and Estela Belén Olivera, from the University of Mendoza, our copy editor, are also very appreciated.

This work was funded by the National Council of Scientific Research of Argentina (CONICET) and has been performed partially under the Project REINFORCE (GA 872859) with the support of the EC Research Innovation Action under the H2020 Programme SwafS-2019-1.